\documentclass[aps,prd,superscriptaddress,preprint]{revtex4-1}
\usepackage{amsmath,amssymb,bm}
\usepackage{graphicx,graphics,color}
\usepackage[colorlinks=true,linkcolor=green,bookmarks=true]{hyperref}
\usepackage{epsfig}
\usepackage{latexsym}
\usepackage{rotating}
\usepackage{subfigure}
\usepackage{centernot}
\usepackage{longtable}
\usepackage{hhline,multirow,tabularx}  % for nicer tables
\usepackage[utf8]{inputenc}
\usepackage{newunicodechar}
\newunicodechar{，}{,}
\usepackage{float}
\usepackage{mathrsfs}
\usepackage{xr}
\usepackage{adjustbox}
\usepackage{verbatim}

\newcommand{\RNum}[1]{\uppercase\expandafter{\romannumeral #1\relax}}

\newcommand{\cev}[1]{\reflectbox{\ensuremath{\vec{\reflectbox{\ensuremath{#1}}}}}}
\begin{document}

\title{Nucleon axial form factors in generalized nonlocal chiral effective theory  }

\author{Y.~Salamu}
\affiliation{\mbox{School of Physics and Electrical Engineering, Kashi University, Kashgar, 844000, Xinjiang, China}}

\author{Haximjan Abdusattar}
\affiliation{\mbox{School of Physics and Electrical Engineering, Kashi University, Kashgar, 844000, Xinjiang, China}}

\vskip 1 cm

\begin{abstract}
 In this work, we first investigate the chiral transformation properties of the 
nonlocal pion field operator and construct a generalized nonlocal chiral 
Lagrangian that is invariant under $\mathrm{SU}(2)$ chiral symmetry 
transformations. As a simple application, we then calculate the leading and next‑leading order pion one loop corrections to the nucleon axial form 
factors within nonlocal chiral perturbation theory. Then, we fit the 
next leading order  and next‑next leading    order low energy coupling  constants (LECs) to lattice QCD 
data. The fitted results show  that the nonlocal LECs are comparatively 
smaller than their local counterparts. Finally, using these LECs, we compute 
the $Q^{2}$ dependence of the nucleon axial form factors. The numerical 
results indicate that the nonlocal nucleon axial form factors are consistent with 
lattice QCD data in a wide $\rm Q^{2}$ range, up to 
$\rm Q^{2} = 1~\mathrm{GeV}^{2}$.
\end{abstract}

\date{\today}
\maketitle
\section{ Introduction  }
\label{sec:1}
One of the main issues in  quantum field theories (QFT) is that, in  most cases, the loop corrections to the physical observables are  ultraviolet(UV) divergent. In renormalizable quantum field theories, divergences can be removed by introducing corresponding counterterms for  these parameters such as mass, charge, and coupling constants. To renormalize a given theory, firstly, it is necessary to replace the original loop integral with a corresponding convergent one by modifying the integral limit, dimension,or the propagator and then calculate the loop integral. Such a mathematical process is  usually  referred to as regularization. Traditionally, dimensional (DR) and  Pauli-Villars (P-V) regularization   methods  have been widely  applied to regularize  the  divergence of the theories. However, Pauli-Villars regularization technique  violates the symmetry of theories, i.e, the Lorentz invariance, gauge symmetry and unitarity  \cite{Muta:1998vi}. Therefore, dimensional regularization has been more preferred in QFT since it preserves both the Lorentz and gauge invariance of theories. Nevertheless, the dimensional regularization cannot be applied to string field theory owing to nonlocal properties of strings \cite{Evens:1990wf,Buoninfante:2018lnh,Moffat:1990jj,Basu:2001rh}. In this case, nonlocal regularization has been introduced to cure the UV infinities of string loop integrals and extended to the gravity and other aspects of  field theory\cite{Efimov:1967pjn,Alebastrov:1972dxn,Tomboulis:2015gfa,Kleppe:1991rv,Boos:2020ccj,Mazumdar:2018xjz,Ohta:1990fi,Terning:1991yt,Anikin:1995cf}. Compared with other conventional regularization methods, the main advantage of nonlocal regularization is that it can be implemented in the Lagrangian level. More specifically, one can  obtain a UV finite  loop result by using the nonlocal propagators or  vertices, which can be systematically   generated from the nonlocal Lagrangian in a more self-consistent way. More importantly, the nonlocal Lagrangian  preserves the symmetry of the underlying  theory.

 In fact, the UV divergence stems from the problematic nature of interaction between the  particles,  in which it was assumed that particles  interact with each other at the same space-time point\cite{DiGrezia:2008am}. However, for the composite particles such as baryon and meson, this assumption breaks down  due  to the non-negligible finite size of particles in question. Therefore it is necessary to explore a fundamental  theory that is  compatible with the non-point  nature of particle. Conventionally, as an important theoretical tool of describing   the interactions of the baryons and mesons, the  basic framework of chiral effective field theory and its renormalization have been extensively investigated.  Although chiral perturbation theory successfully explained the  experimental data for the nucleon magnetic moments  and effective charge radii, the $Q^2$-dependence of electromagnetic and axial vector  form factors in the large $Q^2$ region is incompatible with data\cite{Fuchs:2003ir,Kubis:2000zd,Schindler:2006it}. On the other hand, in Refs\cite{Ohta:1990fi,Holdom:1989jb,Salamu1,Salamu2}, some  preliminary studies have been   performed and constructed  the basic theoretical framework of nonlocal regularization of electromagnetic interaction. In addition to this, in Ref \cite{Ivanov:1996pz,Anikin:1995cf,He1,He2,He3,He4,Wang03}, it was shown that the nonlocally regularized pion and nucleon electromagnetic form factors successfully describe the data  in the  $\rm 0 \le Q^2  \le1G{\rm{e}}V^2 $  region. This intrigued the interest of exploring the  feasibility of integrating the    nonlocal regularization with chiral perturbation theory. Although there are some preliminary  studies on nonlocal renormalization of chiral meson loop corrections to electromagnetic form factors (EMFFs) and parton distribution functions (PDFs)  which are only  $U(1)$ gauge invariant, the systematic nonlocal renormlization of chiral effective theory is not yet  well understood  especially one considers the full  $\rm SU(2)$ or $\rm SU(3)$  chiral symmetry of nonlocal meson field. More importantly, although the $Q^2$-dependence of nucleon electromagnetic form factors in the large  momentum transfer region was improved by the   nonlocal framework, the nonlocal effects to the nucleon axial structure are not well defined. This prompted our interest to investigate  the nucleon axial form factors in the nonlocal framework.  
 
 In this paper, we will develop  a  more  generalized  nonlocal  regularization method for chiral meson loops in a more systematic way. Just like dimensional renormalization, the meson loop results from such nonlocal Lagrangian are ultraviolet finite and Lorentz invariant. More importantly, in contrast to  previous formulations of nonlocal chiral theory, this  method appears to be more efficient when we take into account the nonlocal interaction between the nucleon axial vector current and axial vector gauge field. Since, the newly constructed nonlocal Lagrangian is invariant under $\rm SU(2)$  chiral transformation. In addition, the generalized nonlocal chiral Lagrangian systematically yields regularized meson loop contributions at all orders without manually  modifying  the propagator or loop integral  as in  Pauli-Villars and sharp cutoff regularization.

This paper is  organized as follows: In Sec.\ref{sec:2} we  will shortly review the local chiral transformation properties  of  meson and fermion fields  and  analyze  the chiral transformation of nonlocal meson fields and explicit  parametrization of  gauge link operators, then construct nonlocal chiral  Lagrangians  for pion-nucleon interactions . In  Sec.\ref{sec:3}, to explicitly illustrate the nonlocal effect on  nucleon axial form factors, we calculate pion one-loop corrections to   the  nucleon axial form factors within the nonlocal framework. In  Sec.\ref{sec:4}, we first fit low energy coupling constants (LECs) to the Lattice data, then numerically  compute  the $Q^2$-dependence of axial form factors and discuss  the axial charge as well as axial mean square radii.
\section{Nonlocal Chiral effective Lagrangian }
\label{sec:2}
In this section, we develop a more general nonlocal regularization scheme for  chiral meson loops. As mentioned  in Sec.\ref{sec:1}, nonlocal regularization of ultraviolet  divergence of pion loops  can be encoded into the  Lagrangian without violating the underlying  symmetry of the theories. The key step  to obtain a nonlocally regularized Lagrangian   is that, first of all, delocalize the field that is  associated with  loop contributions, then construct a nonlocal Lagrangian by  replacing the local field operator with its corresponding delocalized version and corresponding gauge link operators, which guarantees the chiral symmetry  of the  nonlocal Lagrangian. Actually, such a procedure can be implemented by delocalizing  the field operators in both kinetic and  interaction terms which can regularize the ultraviolet divergences of loop contributions in the same manner. Therefore, we have choice of  where to locate the delocalized field operator in the Lagrangian. According to the location of the nonlocal field operator, the nonlocally regularized  Lagrangian can be constructed in the inhomogeneous and homogeneous fashion. In  the inhomogeneous nonlocal field  theory, the delocalized field operator  is located in the interaction  Lagrangian while the kinetic term keeps its local form, and they can be transformed into each other  by redefining the field operator\cite{Buoninfante:2018mre}. In other words, in the inhomogeneous nonlocal theory the propagator for the  field that is associated with loop corrections  is the same as the local one and the interaction vertex introduces an extra regulator  function which increases the superficial dimension of denominator of  loop integral. Consequently,  the loop result is UV convergent. However, the price for this is that the symmetry of  the  theory  is  violated. To restore symmetry, it is necessary to introduce corresponding gauge link operators based  on the  symmetry transformation of nonlocal field. To  this end, we first will review the $\rm SU(2)$ chiral  symmetry  of  local  chiral perturbation theory under the investigation. For the local pion-nucleon  interactions, as we have learned, it is governed by the chiral Lagrangians which fulfill the $\rm SU(2)$  chiral symmetry. Generally, fundamental components of $\rm SU(2)$ chiral perturbation theory are the nucleon and pion fields. Under chiral transformation, they transform as,
  \begin{equation}
     U'(x)= R(x)  U(x)  L^\dag (x),  
     N '(x)=K(x)N (x),
    \label{eq:1}
  \end{equation}
 where $ U(x)$ is the exponential representation of pion  and is defined as $U(x)= {\rm Exp} [ i \frac {\phi(x)}{f} ]$  with the $\rm SU(2)$ pion field  matrix $\phi(x)$, $N (x)$ denotes  nucleon fields, $R(x)$ and $L(x)$ denote the elements of $\rm SU(2)$ chiral group and are parametrized in terms of exponential representations $R(x)={\rm exp} [-i\theta_R(x)]$ and $L(x)= {\rm exp}[-i\theta_L(x)]$,  where matrices  $\theta_R(x)$ and  $\theta_L(x)$ are defined as  $\theta_R(x)=\sum\limits_a \frac{1}{2}\theta_R^a(x)\tau^a$ and $\theta_L(x)=\sum\limits_a\frac{1}{2}\theta_L^a(x)\tau^a$, with the $\tau^a (a=1,2,3)$ denoting  the Pauli matrices, $K(x)$ denotes the  element of $\rm SU_V(2)$.
  
  Our investigation now proceeds to examine  the  delocalization of pion  fields and their  chiral transformations. In fact, a nonlocal   pion  field  operator can readily  be obtained by  replacing the local meson  field operator $\phi(x)$ with the corresponding nonlocal  field operator $\int d^4a H(a)\phi(x+a)$. Nevertheless, the chiral transformation rule for  the nonlocal  pion field $ \int d^4a H(a)  \phi(x+a)$ cannot be well defined according to chiral transformations in Eq.(\ref{eq:1}). In fact, the chiral transformations include vector and axial vector transformations  of pion and nucleon, since  $\rm SU(2)_L \times SU(2)_R \cong SU(2)_V \times SU(2)_A$. For the sake of convenience, we may initially  consider  the coupling between the external  vector gauge  field and the  chiral vector current, where the   local meson field $\rm U(x)$ and nucleon   field obey explicit  $\rm SU(2)_V $ transformation rules. Thus one is able to construct $\rm SU(2)_V $ invariant   nonlocal  Lagrangian by including the nonlocal pion field into  theory and introducing the corresponding  gauge link operator. However, such a method  is only valid when we construct a $\rm SU(2)_V $ invariant nonlocal Lagrangian. As for axial vector couplings, since the local meson field lacks an explicit $\rm SU(2)_A $ transformation behavior so that  one cannot properly define the symmetry properties of nonlocal interaction between  external axial vector field and axial vector current. Therefore, we investigate a more general method in which the nonlocal pion field exhibits approximate ${\rm SU(2)}$ chiral transformation behavior just like its local counterpart. For that purpose, we define   a  nonlocal pion  field operator as,
\begin{equation}
{\rm \hat   U(x)} \equiv  { \rm Exp} \{   \frac{1}{c_0}\int d^4a H(a) {\rm Log}[{\rm G}_R(x,x+a){\rm U } (x+a){\rm G}^{\dag}_L(x,x+a)] \},
\label{eq:2}
\end{equation}
where  $c_0$ is a normalization constant and satisfies  $c_0\equiv\int da  H(a)$, $G_{L(R)}(x,x+a)$ are the gauge link operators, $ H(a)$ is the   meson  regulator function in space-time, ${\rm G}_R(x,x+a)$ and ${\rm G}_L(x,x+a)$ are right and left handed  the gauge link operators. The  regulator function  $\rm H(a)$ has some properties and  satisfies  certain  constraints  which, as discussed  in more  detail in  \cite{Efimov:1967pjn}. It is obvious  that, according to Eq.~(\ref{eq:1}), the  pion field  $\rm U(x+a)$ transforms as,
\begin{equation}
\begin{split}
&{\rm U}'(x+a)={\rm R}(x+a){\rm U}(x+a){\rm L}^\dag(x+a).
\end{split}
\label{eq:3}
\end{equation}
Indeed, such a transformation gives rise to chiral phase factor at two different space-time points. To cancel this, the gauge link operators have to transform,
\begin{equation}
\begin{split}
G'_{L[R]}(x,y)= L(x)[ R(x)]G_{ L[R]}(x,y){ L}^\dag(y)[ R^+(y)].
\end{split}
\label{eq:44}
\end{equation}
Using these transformation properties in Eqs.~(\ref{eq:3}) and (\ref{eq:44}), the  nonlocal pion field operator ${\rm \hat   U(x)}$ then  transforms as,
\begin{equation}
{\rm \hat   U'(x)}= {\rm Exp}\{\frac{1}{c_0} \int da {\rm  H(a)} {\rm Log}[{\rm R(x)}{\rm G}_R(x,x+a){\rm U } (x+a){\rm G}^\dag_L(x,x+a){\rm L^\dag(x)}]\}.
\label{eq:4}
\end{equation}
Notice that left and right-handed chiral phase factors $\rm L(x)$ and $\rm R(x)$  are now in the same space-time point. This allows   us to  define explicit chiral transformations of nonlocal field operator ${\rm \hat   U(x)}$. Actually, expression in Eq.(\ref{eq:4}) can be further simplified using  Baker-Campbell-Hausdorff formula, which states that product of two matrix exponential $ e^Xe^Y$ can be expressed as $Z={\rm Log}( e^Xe^Y)$ and $Z$ is given by in terms of  series of commutator  of  $X$  and $Y$ up to third  order as,
  \begin{equation}
    Z=X+Y+\frac{1}{2}[X,Y]+\frac{1}{12}[X,[X,Y]]-\frac{1}{12}[Y,[X,Y]]+\cdot\cdot\cdot
  \label{eq:5}
  \end{equation}
On the other hand, the operator  ${\rm G}_L(x,x+a){\rm U } (x+a){\rm G}^+_R(x,x+a)$ is independent of local phase factors and by virtue of  Baker-Campbell Hausdorff formula, it  can be notationally written  as,
 \begin{equation}
 {\rm G}_R(x,x+a){\rm U } (x+a){\rm G}^+_L(x,x+a)\equiv e^{iX(x,a)}.
 \label{eq:22e}
\end{equation}
 where $X(x,a)$ is a nominal  operator  that we no need to know (keep it mind that ${\rm G}_L(x,x+a)$ and ${\rm G}_R(x,x+a)$ will be parameterized in terms of right and left handed external vector  field $l_\mu(x)$ and $r_\mu(x)$ ). Applying Baker-Campbell-Hausdorff formula again,  Eq.(\ref{eq:4}) is written as
\begin{equation}
\begin{split}
{\rm \hat   U'(x)}&= {\rm Exp} \biggr\{ -i\theta_{R}(x)+i\theta_L(x)+\frac {i}{c_0}\int da {\rm  H(a)}X(x,a)+\frac{1}{2c_0}[\theta_L(x)+\theta_R(x),\int da H(a)X(x,a)]\\
&+...\biggr\},
\end{split}
\label{eq:115a}
\end{equation}
where the ellipsis represents the higher order Baker-Campbell Hausdorff series. Neglecting higher order series in the exponent, chiral transformation of nonlocal meson field $\hat {\rm  U}'(x)$  finally can be expressed as,
\begin{equation}
 \hat  {\rm U}'(x) ={\rm R}(x){\rm {\hat   U}}(x){\rm L}^\dag(x).
 \label{eq:7}
\end{equation}
 Obviously, chiral transformation of nonlocal meson field ${\rm \hat  U}(x)$ is the same as its  local counterpart ${\rm   U}(x)$. Furthermore, in the local limit  $\rm H(a)=c_0\delta^4(a)$  nonlocal pion exactly recovers   to the  local case $\hat  U(x)= U(x)$. In addition, taking $\phi(x+a)=0$, the  ground state of $\hat {\rm U}$ is $\hat {\rm U}=1$ which is gauge  invariant under $\rm SU(2)_V $ transformation ( for gauge link operators, take $r_\mu=l_\mu=v_\mu $). Finally, these right and left handed  gauge link operators can  easily be parameterized in terms of  left and right handed external vector field by analogy with the Wilson line operators in QCD  \cite{Cheng:1986hu},
\begin{equation}
  G_R(x,{x+a})={\cal{P}} {\rm Exp}[-i  \int_{x}^{x+a} dz^\mu r_\mu (z)],G_L(x,{x+a})={\cal{P}}{\rm Exp}[-i \int_{x}^{x+a} dz^\mu l_\mu (z)],
 \label{eq:8}
\end{equation}
where $\cal{P}$ denotes the  path ordering operator,  $r_\mu (x)$ and $l_\mu (x)$ are the  right and left handed  gauge fields and whose chiral transformation properties are defined as,
\begin{equation}
  l'_\mu (x)= L(x)   l_\mu (x) L^\dag(x)+i L(x) \partial _\mu    L^\dag (x),\quad
   r'_\mu (x)=R  r_\mu (x)R^\dag +iR\partial _\mu R^\dag.
  \label{eq:4l1}
\end{equation}
Now let us consider the strategy of constructing a nonlocal effective Lagrangian for the pion-nucleon interaction. As shown in Eq.(\ref{eq:7}), with the help of the gauge link operators in  Eq.(\ref{eq:8}), the nonlocal pion  field $\rm \hat  U(x)$ exhibits chiral transformation behavior as its local counterpart $\rm  U(x)$. Thus, it is easy to obtain any order nonlocal chiral Lagrangian which will produce nonlocally regularized  chiral meson loops. For example, the  the leading order, next leading order, and next-next leading‑order   $\rm SU(2)$ chiral Lagrangians for nonlocal nucleon-pion interactions are just given by replacing   the local meson field operator $\rm   U(x)$ with the nonlocal one $\rm  \hat U(x)$ in local  Lagrangian\cite{Fettes:2000gb,Fettes:1998ud},
\begin{equation}
\begin{split}
{\cal L}^{\rm (1)}_{\pi N}&= \bar{N}  \bigg\{i \centernot D - m  + {g_A \over 2}  \gamma ^\mu \gamma_5 \hat u_\mu  \bigg\}  N,\\
{\cal L}^{\rm (2)}_{\pi N}&= \bar{N} \bigg\{ c_1\,\langle \chi_+ \rangle
-\frac{c_2}{8m^2} \Big(\langle \hat{u}_\mu \hat{u}_\nu \rangle \{\vec D^\mu,\vec D^\nu\} +\{ \cev {D}^\mu , \cev {D}^\nu  \}  \Big)
+ \frac{c_3}{2} \langle \hat u_\mu \hat u^\mu  \rangle+ \frac{i c_4}{4} \sigma^{\mu\nu} [\hat u_\mu,\hat u_\nu]\bigg\} N,\\
\mathcal{L}_{\pi N}^{(3)}&=\bar{N}\big\{\frac{d_{16}}{2}\gamma^\mu\gamma_5\langle\chi_+\rangle \hat {u}_\mu +\frac{d_{22}}{2}\gamma^\mu\gamma_5[D_\nu,F_{\mu\nu}^-]\big\}N+...
\label{eq:eff2}
\end{split}
\end{equation}
where $N$  denotes the  nucleon field, which includes proton and neutron via the  definition  $N=(p,n)^T$ , $m$ is the nucleon mass in the chiral limit, $g_A$, $c_i(i=1...4)$, $d_{16}$ and $d_{22}$ are the leading, next‑to‑leading‑order, and next‑to‑next‑to‑leading‑order  low energy coupling constants in  the chiral limit, the covariant derivative $D_\mu$ and the nonlocal vector and axial vector connections $\hat \Gamma_\mu$ and $\hat u_\mu$ are defined as,
\begin{eqnarray}
D_\mu&=&\partial_\mu+\hat{\Gamma}_\mu,\nonumber\\
 \hat \Gamma_\mu &=&\frac {1}{2} [ {\hat u}^{\dag}  (\partial_\mu-ir_\mu  )\hat{ u}+\hat   u (\partial_\mu-il_\mu ){\hat  u}^{\dag} ],\nonumber\\
 \hat u_\mu&=& i[{\hat u}^{\dag}  (\partial_\mu-ir_\mu  )\hat{ u}-\hat   u (\partial_\mu-il_\mu ){\hat  u}^{\dag}] ,\nonumber\\
 \chi_{\pm}&=& {\hat u}^\dag\chi  {\hat u}^\dag \pm  {\hat u} \chi^\dag  {\hat u}, \nonumber\\
 F^{\pm}_{\mu\nu}&=& {\hat u}^\dag F^R_{\mu\nu}  {\hat u}\pm {\hat u} F^L_{\mu\nu} {\hat u}^\dag,\nonumber\\
 F_{\mu\nu}^{L}&=&\quad\partial_{\mu}l_{\nu}-\partial_{\nu}l_{\mu}-i\left[l_{\mu},l_{\nu}\right],\nonumber\\
 F_{\mu\nu}^{R}&=&\quad\partial_{\mu}r_{\nu}-\partial_{\nu}r_{\mu}-i\left[r_{\mu},r_{\nu}\right],
\label{eq:12}
\end{eqnarray}
where the nonlocal $\hat u$ is defined as  $\hat u=\sqrt{\hat{\rm U}}$, $\hat u^\dag=\sqrt{\hat{\rm U}^\dag}$, $\chi $  symbolically  represents the explicit chiral symmetry breaking term and is defined in terms of  liner combination of external scalar and pseudoscalar field $\chi =2B_0[s(x)+ip(x)]$. According to the chiral transformation property of nonlocal meson field  $\rm  \hat U(x)$, the nonlocal $\hat u $ transforms as  $\hat u'=R \hat uK^\dag =K\hat uL^\dag $. Under this notation, nonlocal gauge connections $\hat \Gamma_\mu$ and $\hat u_\mu$ transform as,
\begin{eqnarray}
 \hat \Gamma'_\mu &=&K\hat \Gamma_\mu K^\dag+ iK^\dag \partial_\mu K,\nonumber\\
 \hat u'_\mu&=&K\hat u_\mu K^\dag.
\end{eqnarray}
Employing  the generalized  $\rm SU(2)$ nonlocal Lagrangians, as an example,  it is necessary  to discuss the consistency between two results from  the previous  $\rm U_V(1)$ invariant  nonlocal  pion-nucleon interactions and  $\rm SU(2)$   nonlocal chiral Lagrangians in  Eq.~(\ref{eq:eff2}). To be specific, expanding Eq.~(\ref{eq:eff2}) in terms of pion and gauge fields gives rise to pion-nucleon interactions as,
\begin{eqnarray}
  {\cal L}^{\rm (1)}_{\pi N}&=& -\frac{g_A}{2f_\phi c_0}\int d^4a H(a)\bar N(x) \gamma ^\mu \gamma^5 \partial_\mu\phi(x+a)  N(x)+\frac{i}{8f^2_\phi c_0^2}\int d^4a_1 H(a_1)\int d^4a_2 H(a_2)\nonumber\\
 && \bar N(x) \gamma ^\mu[ \phi(x+a_1) \partial_\mu\phi(x+a_1)- \partial_\mu\phi(x+a_1) \phi(x+a_2) ] N(x)+\bar  N(x) \gamma ^\mu V_\mu (x) N(x)\nonumber\\ &&+i\frac{g_A}{2 f_\phi c_0}\int da H(a)\bar N(x)\gamma ^\mu \gamma ^5[V_\mu (x+a)\phi (x+a)-\phi (x+a)V_\mu (x+a)]N(x) \nonumber\\
  &&+i\frac{g_A}{2 f_\phi c_0}\int da H(a) \bar N(x)\gamma ^\mu \gamma ^5[\partial_\mu\phi (x+a)\int^x_{x+a} dz^\nu V_\nu (z)-\int^x_{x+a} dz^\nu V_\nu (z)\partial_\mu\phi (x+a)]N(x)\nonumber\\
  &&+\frac{g_A}{c_0}\int da H(a) \bar N(x)\gamma ^\mu \gamma ^5 A_\mu (x+a)N(x)+i\frac{1}{4f_\phi c_0^2}\int da_1 H(a_1)\int da_2 H(a_2)\bar N(x)\gamma ^\mu\nonumber\\
  &&\biggl\{[A_\mu (x+a_2)+A_\mu (x)]\phi (x+a_1)-\phi(x+a_1)[A_\mu (x+a_2)+A_\mu (x)]\biggr\}N(x)\nonumber\\
  &&+i\frac{1}{4f_\phi c_0^2}\int da_1 H(a_1)\int da_2 H(a_2)\bar N(x)\gamma ^\mu \biggl\{\int^x_{x+a_2} dz^\nu A_\nu (z)\partial_\mu\phi (x+a_1)\nonumber\\
  &&-\partial_\mu\phi (x+a_1)\int^x_{x+a_2} dz^\nu A_\nu (z)\biggr\}N(x)+...
 \label{eq:14}
 \end{eqnarray}
where ellipsis denotes  higher-order pion derivative terms. From Eq.(\ref{eq:14}) we can see that  interactions which contains the vector gauge field $V(x)$  is identical  to the previous nonlocal result that was constructed  based on the  $\rm U_V(1)$ vector gauge symmetry of nonlocal meson-nucleon interactions   \cite{Salamu1,He1}. 
In addition, it is worth mentioning  that the pion  field and regulator function $H(a)$ are in  one-to-one correspondence. In other words,  every pion  field can be regularized by a corresponding regulator function so that pion  loop integrals are convergent at any chiral order. More importantly, in generalized  $\rm SU(2)$ nonlocal framework, the  nucleon axial vector current couples to a nonlocal axial vector gauge field. Consequently, the nucleon axial form factor at the tree level is proportional to the nonlocal regulator function, which has a non-negligible effect on  the physical observables such as nucleon axial charge and axial mean square radii. 
\section{Nucleon AXIAL FORM Factors }
\label{sec:3}
\subsection{definition of axial form factors }
Theoretically, by analogy with  electromagnetic form factors, the matrix elements of nucleon  isovector axial vector currents can be parameterized in terms of nucleon axial form factors as\cite{Schindler:2006it,Bernard:1994wn}, 
\begin{equation}
    \langle N(p',s') \vert A^{\mu,a}\vert N(p,s) \rangle = \bar{u}(p')
     \bigg[\gamma^\mu G_A(q^2)+ \frac{q_\mu}{2 m} G_P(q^2)\bigg]\frac{\tau^a}{2} \gamma_5 u(p),
\end{equation}
where $u(p)$ and $u(p')$ are  the nucleon spinors for initial and final states, $\tau^a$ are  the Pauli isospin matrices, $q$ is the  four momentum transfer $q=p'-p$ and  $Q^2=-q^2$, scalar functions $G_A(q^2)$ and $G_P(q^2)$ denote the nucleon axial-vector and induced pseudoscalar form factors. To describe  the axial charge  distribution, it is necessary  to introduce the mean square of axial charge radii, which is defined by the slope of $G_A(Q^2)$ at zero momentum transfer. To this end, the nucleon axial vector  form factor  $ G_A(Q^2) $  can be expanded around  the small $Q^2$ as, 
\begin{equation}
    G_A(Q^2)=g_{A}[1-\frac{1}{6}\left\langle {r_A^2 } \right \rangle Q^2+...],
\end{equation}
where $ g_{A}$ and $\left\langle {r_A^2 } \right \rangle$ are the renormalized  nucleon axial charge and squared axial charge radii, respectively. Furthermore, the nucleon induced pseudoscalar coupling constant is defined as, 
\begin{equation}
    g_P\equiv \frac{m_\mu}{2m_N}G_P(0.88m^2_\mu),
\label{eq:36}
\end{equation}
where $m_\mu$  denotes  the Muon mass. Experimentally, induced pseudoscalar coupling constant $g_P$ is determined from the muon-Hydrogen capture experiment.
\subsection{One loop contributions to the nucleon self energy }
\begin{figure*}
    \centering
    \includegraphics[width=0.5\textwidth]{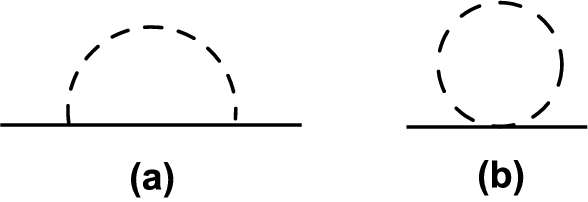}
    \caption{Leading and next- to-leading  order pion  one loop corrections to the  nucleon self  energy, dashed and solid lines denote  the pion and nucleon fields, respectively.}
    \label{fig:loop10}
\end{figure*}

To obtain a renormalized  the axial form factors $G_A(q^2)$ and $G_P(q^2)$, the loop results  should be multiplied by the wave function renormalization constant $Z$. Generally, wave function renormalization constant $Z$ and nucleon physical mass $m_N$ can be obtained from  the corresponding  nucleon self-energy according to the definition,
\begin{eqnarray}
       m_N&=&m-4m_\pi^2c_1+\Sigma(\not p,m),\\
    Z&=&1+\frac{d \Sigma(\not p,m)}{d \not p}|_{\not p=m},
\label{eq:mass}
\end{eqnarray}
where the nucleon self energy $\Sigma(\not p,m)$ includes leading and next-to- leading order pion one loop  contributions in Fig.~\ref{fig:loop10} and they explicitly can be expressed as,  
\begin{eqnarray}
%\begin{equation}
%\begin{split}
 \Sigma_{a}(\not p,m)&=& i \, \frac{3g^2_A}{4f^2_\phi}\int {\frac{{{d^4}k}}{{{{\left( {2\pi } \right)}^4}}}} \frac{{\not k {\gamma _5} (\not p - \not k + m) \not k \gamma _5}}{{{D_\phi }(k){D_N}(p - k)}}\tilde{F}^2(k), \nonumber\\
 \Sigma_{b}(\not p,m)&=&-i \, \frac{1 }{f^2_\phi} \int {\frac{{{d^4}k}}{{{{\left( {2\pi } \right)}^4}}}} \frac{-6c_1 m_\pi^2 +3\frac{ \,c_2 }{m^2}(p.k)^2+3 c_3  \,k^2}{ D_\phi (k)}\tilde{F}^2(k),
%\end{split}
\label{eq:selfenenrgy}
%\end{equation}
\end{eqnarray}
where $\tilde{F}(k)$ is Fourier transform of the regulator  function $H(a)$ and is defined as  $\tilde{F}(k)=\frac{1}{c_0}\int d^4aH(a)e^{-ik.a}$, the $D_N(k)$ and $D_\pi(k)$ are  defined as 
$D_\pi(k)=k^2-m_\pi^2+i\epsilon$ and $D_N(k)=k^2-m^2+i\epsilon$
\subsection{One loop contributions to the  axial form factors }
\begin{figure*}
    \centering
    \includegraphics[width=0.8\textwidth]{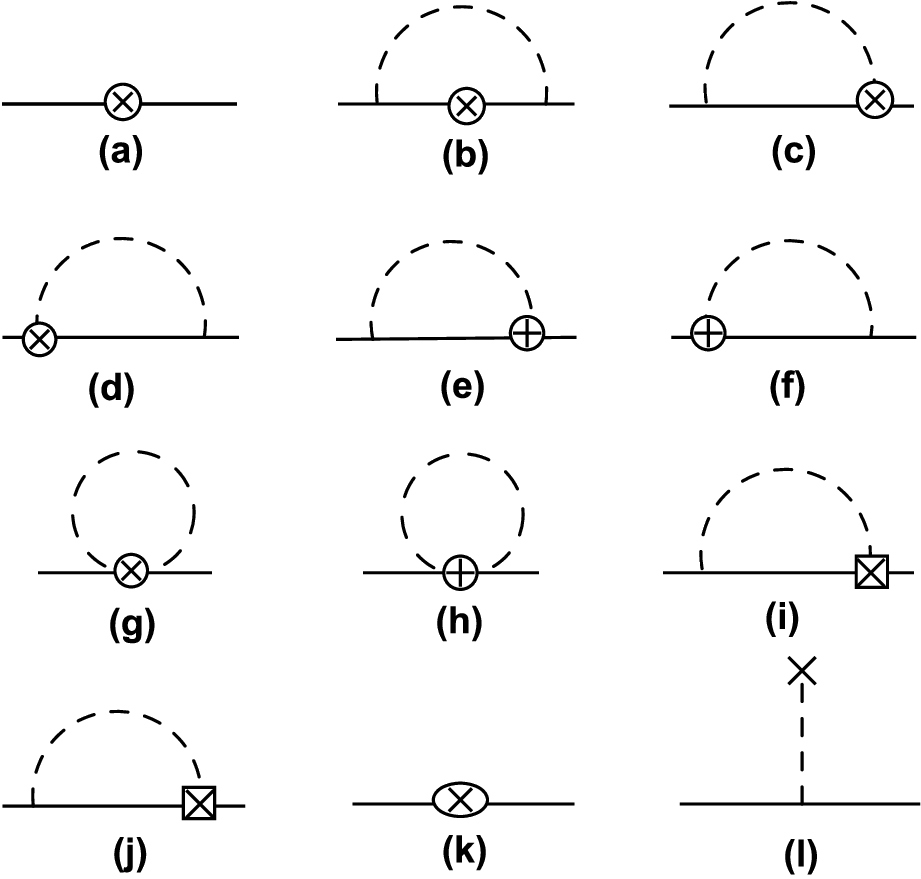}
    \caption{Pion  one loop corrections to the  nucleon axial form factors are shown, where the crossed circle, the crossed square and crossed ellipse  represent the leading, next-to-leading and next-next-to  leading order   axial vector currents, the circled plus is next leading order additional  gauge link axial vector current, and the cross with a dashed line is the pion pole dominant diagram.}
    \label{fig:loop1}
\end{figure*}

In this subsection, we analytically calculate   the  pion one loop corrections to the nucleon axial form factors within the nonlocal framework. Compared to the local case,  as shown in Eq.~(\ref{eq:14}), the tree level coupling between the  nucleon axial vector current  and axial vector gauge field is proportional to a nonlocal regulator. Consequently, such a factor   modifies  the shape of $Q^2$-distribution as well as the mean square radii even at tree-level. Therefore, as a simple example, 
 we explore the effects of non-locality on  the distribution of  nucleon axial form factors.  Using  the nonlocal interactions and axial vector currents , in  Fig.~\ref{fig:loop1} we  display  down relevant leading order, next-to-leading -order and next-next-to leading order  Feynman diagrams that will contribute to the  nucleon axial form factors. From   Fig.~\ref{fig:loop1}(a) one can easily write down the axial vertex  operator for the  tree diagram as, 
 \begin{equation}
    \Gamma_a^\mu(p,q)=g_A \tilde F(q)\bar{u}(p')
\gamma^\mu  \gamma_5 u(p).
\label{eq:26}
\end{equation}
 From   Eq.~(\ref{eq:26}) we can obtain that at tree level, the nucleon axial vector form factor is given by $G_{A,\rm tree}(q^2)=g_A \tilde F(q)$. Notice that in the  nonlocal case the tree level axial form factor explicitly  depends on regulator $\tilde F(q)$. This is totally different from the local case, where $Q^2$-dependence only comes from the loop contributions. Moreover, this is also a direct evidence for the non-point nature of the pion since the factor $\tilde F(q)$  has a nonzero contribution to the pion axial charge radii. In the same way, vertex operator for  Fig. \ref{fig:loop1}(a)  is given by,
\begin{eqnarray}
\Gamma_b^\mu(p,q)&=&-i\frac{g^3_A}{4f^2_\phi}\int\frac{dk^4}{(2\pi)^4}\bar u(p')\not k \gamma^5\frac{\not p'-\not k+m}{D_N(p'-k)}\gamma^\mu\gamma^5 \frac{\not p-\not k+m}{D_N(p-k)}\not k \gamma^5\frac{1}{D_\pi(k)} u(p)\\
&&\tilde F^2(k)\tilde F(q),
\label{eq:27}
\end{eqnarray}
Note that  the contribution of Fig.\ref{fig:loop1}(a), apart from  $\tilde F(q)$,  is proportional to  regulator function     $\tilde F^2(k)$ which arises from the nonlocal pion field. More importantly, the loop  integral is UV  convergent if  $\tilde F^2(k)$  is properly parameterized to increase the superficial dimension of of loop integrand. In the same manner, loop contribution of   Fig. \ref{fig:loop1}(b) is expressed as, 
\begin{eqnarray}
\Gamma_{c+d}^\mu(p,q)&=&-i\frac{g_A}{2f^2_\phi}\int\frac{dk^4}{(2\pi)^4}\bar u(p')[\not k \gamma^5\frac{\not p'-\not k+m}{D_N(p'-k)}\gamma^\mu+\gamma^\mu \frac{\not p-\not k+m}{D_N(p-k)}\not k \gamma^5]\frac{1}{D_\pi(k)} u(p)\nonumber\\
&&\tilde F^2(k)[1+\tilde F(q)].
\label{eq:28}
\end{eqnarray}
As mentioned in the previous section, the nonlocal Lagrangian includes additional  gauge link interactions, which generate additional Feynman  diagrams like    Figs. \ref{fig:loop1}(e) and  (f). Correspondingly, the loop  contribution is given by,  
\begin{eqnarray}
\Gamma_{e+f}^\mu(p,q)&=&i\frac{g_A}{2f^2_\phi}\int\frac{dk^4}{(2\pi)^4}\bar u(p')[\not k \gamma^5\frac{\not p'-\not k+m}{D_N(p'-k)}\not k-\not k \frac{\not p-\not k+m}{D_N(p-k)}\not k \gamma^5]\frac{1}{D_\pi(k)} u(p)\nonumber\\
&&\tilde F^2(k)[1-\tilde F(q)]\frac{1}{q^\mu}
\label{eq:29}
\end{eqnarray}
Obviously, the loop contribution from such a diagram vanishes in the local limit $\tilde F(q)=1$. In addition to this, the tadpole diagram [ Fig.\ref{fig:loop1}(g)]  that arises from  the leading order Lagrangian is given by, 
\begin{equation}
    \Gamma_g^\mu(p,q)=-i\frac{g_A}{3f^2_\phi}\int\frac{dk^4}{(2\pi)^4}\bar u(p')\frac{1}{D_\pi(k)}\gamma^\mu \gamma^5 u(p)\tilde [F^2(k)\tilde F(q)+2F^2(k)].
\label{eq:30}
\end{equation}
Similarly, in the nonlocal case additional gauge link interaction yields a tadpole-like diagram [ Fig. \ref{fig:loop1}(h)] and the corresponding  contribution can be written  as, 
 \begin{equation}
    \Gamma_h^\mu(p,q)=0
\label{eq:31}
\end{equation}
Note  that tadpole-like additional diagram  does not contribute to axial form factors due to the oddity of the loop integrand. As for the  next leading order contribution, it  only includes  Figs. \ref{fig:loop1} (i) and (j) and the loop contributions are  given by,
\begin{eqnarray}
\Gamma_{i+j}^\mu(p,q)&=&i\frac{g_A}{f^2_\phi} \int\frac{dk^4}{(2\pi)^4}\bar u(p')[\not k \gamma^5\frac{\not p'-\not k+m}{D_N(p'-k)}(2c_3k^\mu+c_4[\gamma^\mu,\not k])\nonumber\\
    &&+(2c_3k^\mu-c_4[\gamma^\mu,\not k]) \frac{\not p-\not k+m}{D_N(p-k)}\not k \gamma^5]\frac{1}{D_\pi(k)} u(p)\tilde F^2(k)\tilde F(q),
\label{eq:32}
\end{eqnarray}
where the $c_3$ and $c_4$ are the next leading order effective coupling constants. At next-next leading order,  we only take into account  the tree diagram  Fig.\ref{fig:loop1} (k) and the vertex operator reads, 
\begin{equation}
\Gamma_k^\mu(p,q)=\bar u(p')\gamma^\mu \gamma^5 u(p)(4m_\pi^ 2d_{16}+d_{22}q^2)\tilde F(q)+(-2m d_{22})\bar u(p')q^\mu \gamma^5 u(p)\tilde F(q).
\end{equation}
Obviously, next-next leading order contribution modifies the  $Q^2$-distribution of the nucleon axial form factors as well as the  axial charge radii of the nucleon. Finally, the external axial vector field couples with the  nucleon by a pion intermediate state. Correspondingly, the vertex operator is given by,
\begin{eqnarray}
\Gamma_l^\mu(p,q)=-\frac{2m g_A}{q^2-m^2_\pi} \bar u(p)q^\mu \gamma^5 u(p)\tilde F(q)-\frac{8 d_{16} m m^2_\pi}{q^2-m^2_\pi} \bar u(p)q^\mu \gamma^5 u(p)\tilde F(q).
\label{eq:27}
\end{eqnarray}
 From Eq.(\ref{eq:27}) we can see  that the  pion pole dominant  diagram plays a crucial role in the improvement of  the $Q^2$-dependence of the pseudoscalar form factor. By Collecting each loop contribution listed above, the renormalized nucleon axial vector and axial scalar form factors are obtained  as,
\begin{equation}
\begin{split}
G_A(Q^2)&=Z\sum {G^{(i)}_A(Q^2)},\\
G_P(Q^2)&=Z\sum {G^{(i)}_P(Q^2)},
\end{split}
\label{eq:form}
\end{equation}
where $G^{(i)}_A(Q^2)$ and $G^{(i)}_    P(Q^2)$$(i=a,b,c...)$ represent the loop contribution from  each Feynman diagram.
\section{Numerical results  }
\label{sec:4}
\subsection{fitting of LECs}
\begin{figure}[H]
    \centering
\includegraphics[width=0.5\textwidth]{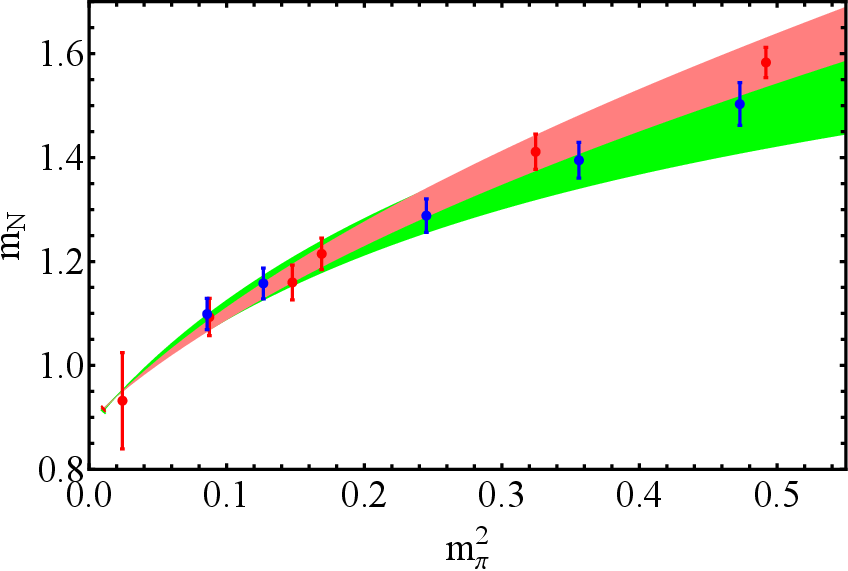}
    \caption{The nucleon mass as a function of pion mass is shown, the red and blue error bars are the lattice data, the green and pink error  bands represent the  theoretical uncertainty that originating   from   uncertainty of $c_1$, $c_2$, $c_{3}$.}
    \label{fig:loop3}
\end{figure}

 To numerically evaluate  the $Q^2$-distribution of the axial form factors and axial charge radii of the nucleon, first of all, we need to parameterize  the  explicit expression of the    regulator function $\tilde F(k)$. There are various   forms of regulator  functions, such as   monopole, dipole and Gaussian types. For convenience's  sake,  in  this work  we  use the dipole form of  regulator function as\cite{Forkel:1994yx,Musolf:1993fu},  
 \begin{equation}
\tilde{F}(k)=\frac{(m_{\pi}^2-\Lambda^2)^2}{(k^2-\Lambda^2+i\epsilon)^2},
\end{equation}
 where the $\Lambda$ represents  the nonlocal  cut-off parameter, which was fitted from the differential cross section for inclusive hadron  production and takes the  value  $\Lambda=(1.0\pm0.1)\rm GeV$\cite{Salamu:2019dok}. In addition, as  input parameters we also have to decide values of next leading and next-next leading order low energy  coupling constants $c_3$ , $c_4$, $d_{16}$ and $d_{22}$. In the standard phenomenological analysis, these parameters can be  extracted from the pion-nucleon scattering. However, in such a method, one has to take into account loop corrections  to the  scattering length. From this point of view,  values of LECs depend on renormalization schemes. Therefore values of locally fitted LECs differ  from those of the nonlocal framework  and  cannot be  applied in this work. If we are  seeking  to extract  them from the  pion-nucleon scattering data within the  nonlocal framework, we have to  handle massive loop contributions. To this end, in a straightforward manner we will refit them to   the lattice data  for the  pion mass dependence of nucleon mass. As seen from Eq.~(\ref{eq:mass}), up to next-next leading  order  nucleon mass $m_N$  proportional to LECs $c_1$ , $c_2$, $c_{3}$. Using the $m_N$ as a fitting  function, we have performed least $\chi^2$ analysis  on two groups of  lattice data  and the  best fit result is  displayed in Fig.~\ref{fig:loop3}, where the error bars  represent the lattice data while the error bands represent the best fit model result. Correspondingly, the best fit values  for $c_1$ , $c_2$, $c_{3}$ are listed  in Tab.\ref{tab:6}. From Tab.\ref{tab:6}, we can observe that best fit values for  next leading order LECs  $c_1$, $c_2$ and $c_{3}$ are  $c_1=(-0.211\pm0.020){\rm GeV^{-1}}$, $c_2 =(-0.221 + 0.021){\rm GeV^{-1}}$, $ c_3=(-0.055\pm 0.084){\rm GeV^{-1}}$ and $c_1=(-0.119\pm 0.036){\rm GeV^{-1}}$, $c_2=(-0.755\pm0.035){\rm GeV^{-1}}$, $c_3=(-0.189\pm0.140){\rm GeV^{-1}}$ with corresponding best fit parameters  $\chi^2/d.o.f=0.18$ and   $\chi^2/d.o.f=0.37$, respectively. It is worth to mentioning  that in the nonlocal case the  LECs are  comparatively smaller than its  local counterpart. In addition to this,  the  nucleon chiral mass $m$ is larger than in the the local case.

\begin{figure}[H]
\centering
\begin{tabular}{ccc}
{\epsfxsize=3.3in\epsfbox{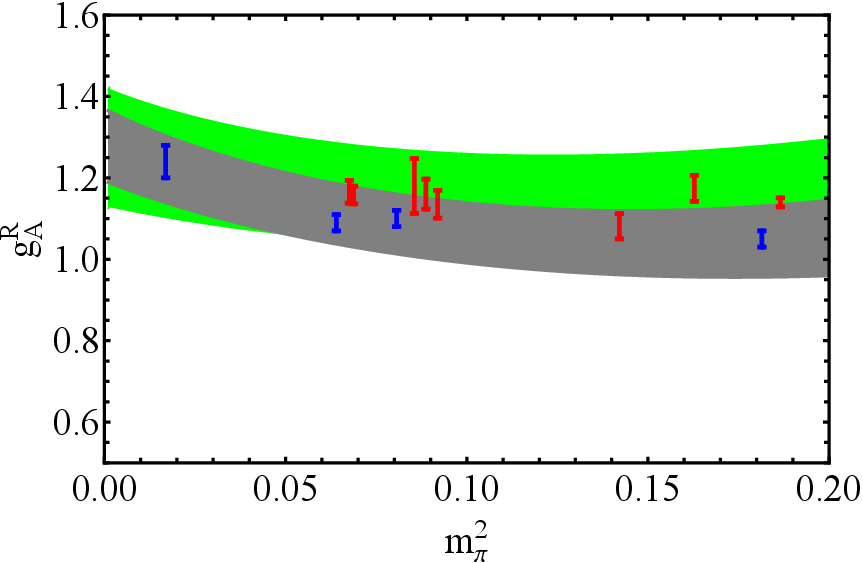}}&
{\epsfxsize=3.5in\epsfbox{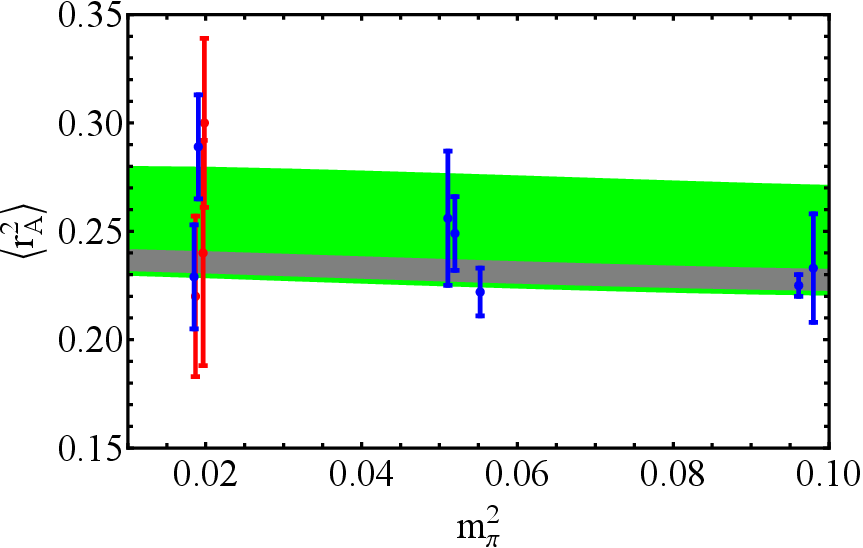}}& 
\end{tabular}
\caption{Nucleon axial charge $g^{R}_{A}$ (left panel ) and squared axial charge radii  $\left\langle {r_A^2 } \right\rangle $  (right panel ) as a function of pion mass, where the red and blue error bars are the lattice data from\cite{Jager:2013kha,Horsley:2013ayv,Edwards:2005ym,Gupta:2017dwj}, the green and pink error  bands represent the  theoretical uncertainty that arising  from   uncertainty of  $c_4$ , $d_{16}$, $d_{22}$.}
\label{fig:55}
\end{figure}

\begin{table}[H]
\begin{center}
\begin{tabular}{c|c|c|c|c|c}
\hline
\hline
$m(\rm GeV)$ &$c_1(\rm GeV^{-1})$ & $c_2(\rm GeV^{-1})$&$c_3(\rm GeV^{-1})$&$\rm \chi^2/d.o.f$& Lattice data  \\
\hline
$1.184\pm0.033$&$-0.211\pm0.020$ & $-0.221 + 0.021$ & $ -0.055\pm 0.084$&0.18&\cite{Walker-Loud:2008rui}\\
\hline
$1.285\pm0.050$&$-0.119\pm 0.036$& $-0.755\pm0.035$& $-0.189\pm0.140$&0.37&\cite{PACS-CS:2008bkb}\\
\hline
\hline
\end{tabular}
\end{center}
\caption{Next leading order LECs $c_1$, $c_2$ and $c_3$ are determined by fitting to the  $m_N$ lattice data from\cite{PACS-CS:2008bkb,Walker-Loud:2008rui}. The nonlocally   renormalized nucleon mass can be obtained from Eq.~(\ref{eq:mass})}
\label{tab:6}
\end{table}
In a similar manner, next and next-next leading order LECs  $c_4$ and $d_{16}$ can be fitted to the lattice data for the  pion mass dependency  of nucleon axial charge $g_A$. The best values of these parameters  are listed  in Tab.\ref{tab:7}. Notice that bare axial charge $g_A$  and  $d_{16}$  are apparently smaller than the local one\cite{Yao:2017fym}. The corresponding fitting curves are displayed  in the left panel of Fig.\ref{fig:55}. Eventually,  the  next-next leading order  coupling constant $d_{22}$ can be fitted to the pion mass dependency of  axial charge radii $\left\langle {r_A^2 } \right\rangle $.  The fitting result shows that best  values of  $d_{22}$ are $d_{22}=(-1.533\pm0.216){\rm GeV^{-2}}$ and $d_{22}=(-1.687\pm0.037){\rm GeV^{-2}}$ with best fit parameters    $\chi^2/\rm {d.o.f}=1.57$ and $\chi^2/\rm {d.o.f}=1.14$.

\begin{table}[H]
\begin{center}
\begin{tabular}{c|c|c|c|c}
\hline
\hline
$g_A$ & $d_{16}(\rm GeV^{-2})$&$c_4(\rm GeV^{-1})$&$\rm \chi^2/d.o.f$& Lattice data \\
\hline
$1.156\pm0.043$ & $1.078\pm0.164$ & $ 0.948\pm	0.219$&1.095&\cite{Edwards:2005ym}\\

\hline
$1.107\pm0.028$& $0.757\pm0.105$& $1.095\pm0.128$&0.735&\cite{Horsley:2013ayv}\\
\hline
\hline
\end{tabular}
\end{center}
\caption{The next and next‑next leading order low‑energy constants  $c_4$ and $d_{16}$ are fitted to the lattice data for  pion-mass dependence of the renormalized axial charge  $g^{R}_A$ of nucleon.}
\label{tab:7}
\end{table}

\subsection{$Q^2$-dependency of axial form factors }
\begin{figure}[H]
\centering
\begin{tabular}{ccc}
\hspace{0.1cm}{\epsfxsize=3.3in\epsfbox{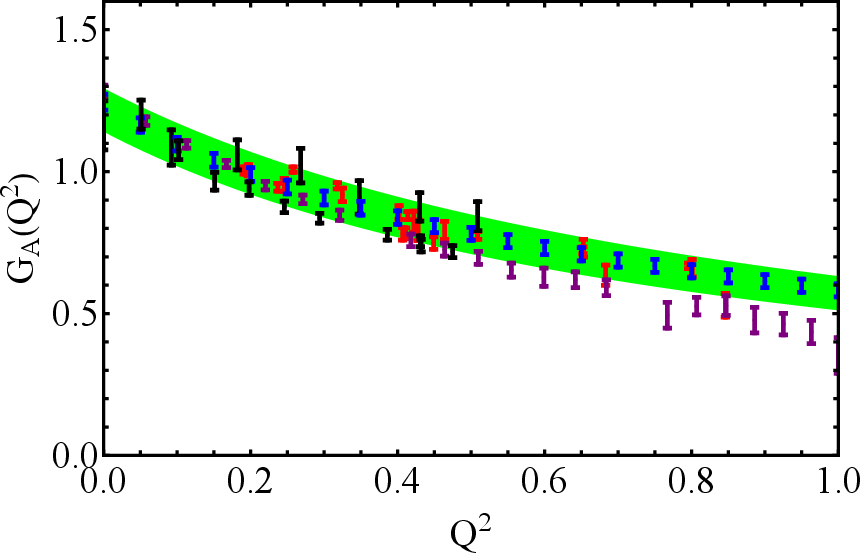}}&
\hspace{0.1cm}{\epsfxsize=3.3in\epsfbox{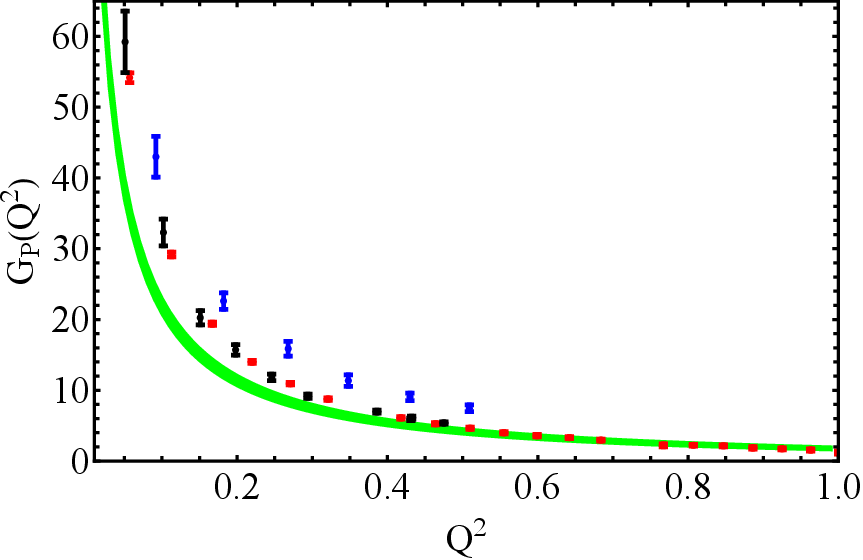}}& 
\end{tabular}
\caption{The $Q^{2}$-dependence  of the nucleon axial form factors 
$G_{A}(Q^{2})$ and $G_{P}(Q^{2})$ is  compared with lattice QCD data from 
Refs.~\cite{Gupta:2017dwj,Alexandrou:2020okk,Capitani:2017qpc,Alexandrou:2023qbg,Jang:2023zts}. 
In the figures, the error bars denote the statistical uncertainties of the 
lattice QCD results, while the green error bands represent the theoretical 
uncertainties arising from the low‑energy constants (LECs) and the regulator 
cutoff $\Lambda$.} 
\label{fig:66}
\end{figure}

Having determined  theses LECs, we are now  in  a position to numerically compute the   $Q^2$-distributions of the  nucleon axial  form factors $G_A(Q^2)$ and  $G_P(Q^2)$. In addition to these parameters, in the numerical analysis  we will use   physical values of pion and nucleon mass as well as the pion decay constant as   $m_\pi=139.57{\rm MeV}$, $m_N=938.92{\rm MeV}$, $f_\pi=92.42{\rm MeV}$. The numerical results for the   $Q^2$-dependence of the nucleon axial vector form factor $G_A(Q^2)$ are  displayed  in  the left panel of  Fig~.\ref{fig:66}. It is obvious that the nonlocal model predictions are in  good agreement with   lattice data in  the wide momentum  transfer    region $0 \le Q^2  \le1{\rm{GeV^2}} $. This implies that the  nonlocal  model significantly  improves the $Q^2$-depencne of axial form factors  compared to the local case, in which the standard (local) chiral perturbation theory can only  describe the experimental data in the region  $0 \le Q^2  \le 0.4{\rm{GeV^2}} $\cite{Schindler:2006it}. In addition to this, the calculations yield a nucleon axial charge of   $g_A=(1.212 \pm0.071)$. Correspondingly, the  mean square radii   of the nucleon axial charge  is obtained as  $\left\langle {r_A^2 } \right \rangle=(0.254\pm0.025){\rm fm^2}$. These values are consistent with recent lattice results and  MuCap experimental value  within  their  uncertainty ranges   
\cite{Capitani:2017qpc,Gupta:2017dwj,Djukanovic:2022wru,Alexandrou:2023qbg,Alexandrou:2020okk}. For the  induced pseudoscalar form factor $G_P(Q^2)$, as shown in right panel of Fig~.\ref{fig:66}, the model result is also consistent with  Lattice data  in the region  $0.3{\rm{GeV^2}} \le Q^2  \le 1{\rm{GeV^2}} $ while in the small-$Q^2$ region, the model result is slightly  lower than lattice data. Consequently, according to the definition in Eq.(\ref{eq:36}), nucleon induced  pseudoscalar coupling  constant is found to be  $g_P=5.308\pm0.195$, which is smaller than the lattice simulation values \cite{Capitani:2017qpc,Gupta:2017dwj,Yamazaki:2009zq,Alexandrou:2023qbg} and  correspond to about $66$\% of the MuCap measurement value    $g_P=8.06\pm0.55$\cite{MuCap:2012lei}. We attribute this discrepancy to neglecting vector‑meson contributions and $d_{18}$-related next‑next leading order effects in the present calculation\cite{Chen:2024kbh,Schindler:2006it}.

\begin{figure}[H]
\centering
\begin{tabular}{ccc}
\hspace{0.1cm}{\epsfxsize=3.3in\epsfbox{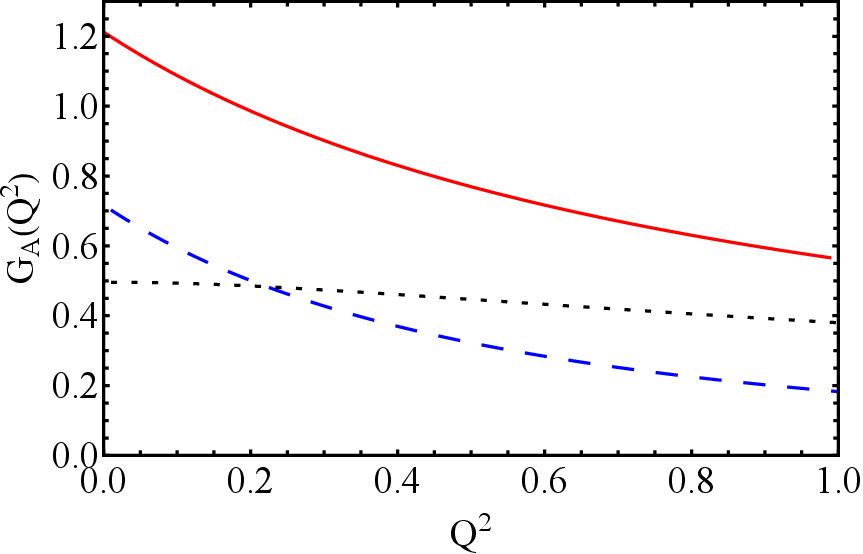}}&
\hspace{0.1cm}{\epsfxsize=3.3in\epsfbox{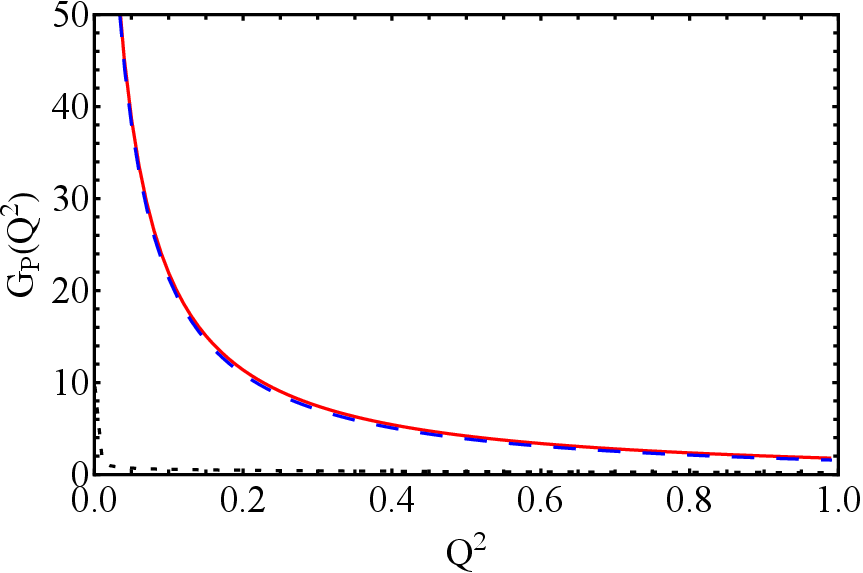}}& 
\end{tabular}
\caption{Tree‑level and loop contributions to the nucleon axial form factor (left panel) and to the induced pseudoscalar form factor (right panel) are shown. The solid red curves denote the total contributions, the blue dashed curves correspond to the tree‑level contributions, and the black dotted curves represent the loop contributions.}
\label{fig:fig60}
\end{figure}

In Fig.\ref{fig:fig60}, we have plotted the  loop and  tree contributions to the axial form factors. As seen from the left panel of  Fig.\ref{fig:fig60}, in the nonlocal case the tree contributions to the $G_A(Q^2)$  are no longer a  constant and strongly improve the  $Q^2$-dependency. This in some sense explains the non-point nature of the nucleon since at tree level the nucleon axial charge radii is not zero. This is contrary to local results in which the axial charge radii is defined  in terms of the slope of loop contribution. As for the pseudoscalar form factors, the pion-pole dominant   diagram accounts for  most of the total contribution, while the loop contribution is substantially suppressed. Such  conclusion  also stands for the  previous result for  nucleon electromagnetic form factors   from   the local chiral perturbation theory\cite{He1}.  

\section{conclusion}
In the conventional field theory, the UV divergence intrinsically is connected with the point particle assumption. On the other  hand, the nonlocal  field theory reflects   the non-point- like nature of particles that need to be investigated. From this viewpoint, nonlocal theory gives rise to  UV convergent loop contributions  by introducing nonlocal field operators and corresponding gauge link operators with  which the underlying symmetry of the theory is still  preserved. In chiral perturbation theory, where the chiral symmetry  serves as a   fundamental symmetry of the underlying theory, the $\rm SU(2)$ or $\rm SU(3)$ chiral symmetry must be  fulfilled when   the nonlocal pion field  is included. 

In this work, we first  investigate  a generalized nonlocal regularization method to regularize the pion loop contribution. Our analysis shows that the novel nonlocal chiral Lagrangians are  still $\rm SU(2)$ chiral invariant and can regularize the   pion loop contributions  up to any chiral order. As an obvious feature   of the  nonlocal chiral interaction，the  coupling between  the nucleon axial current  and the   axial vector field  introduces  additional regulator function  $\tilde F(q)$ which modifies the $Q^2$- dependence of axial form factors and their  slope. Then, as an application of nonlocal regularization, we  have calculated the   pion one loop correction to the  nucleon  axial form factors. To numerically evaluate the axial form factors, we have fit next and next-next leading order  low‑energy coupling constants (LECs) to the  QCD lattice data. The results show that the fitting values of these constants are apparently  smaller than their local counterparts. Finally, using these parameters, we have calculated  the $Q^2$-dependence  of axial form factors $G_A(Q^2)$ and $G_P(Q^2)$. Our analysis shows that the nonlocal model result for axial vector form factor $G_A(Q^2)$ is consistent with   the QCD lattice  data  over a wide momentum transfer  region 
$0 \le Q^2\le1{\rm{GeV^2}} $. Correspondingly, the nucleon axial charge and   mean square radii are obtained   as  $g_A=((1.212 \pm0.071))$,  $\left\langle {r_A^2 } \right \rangle=((0.254\pm0.025){\rm fm^2}){\rm fm^2}$. Moreover, the pseudoscalar form factor $G_P(Q^2)$  is also consistent with  lattice data  within the momentum transfer  region  $0.3{\rm{GeV^2}} \le Q^2  \le 1{\rm{GeV^2}} $. However  in the  small $Q^2$ region the model result is slightly  lower than lattice data. From this, the pseudoscalar coupling constant has been  extracted as $g_P=5.308\pm0.195$.

\section*{Acknowledgments}
This work is supported in part by the National Natural Science Foundation of China under
Grant Nos. 12265016 and 12465012.

\end{document}